\documentclass[prc,twocolumn,showpacs,epsf]{revtex4}
\usepackage{graphicx}
\begin{document}
\draft
\title
{A New  Measurement of the Proton Capture Rate on $^7$Be  and the 
$S_{17}$(0) Factor}
\author{
L. T. Baby$^1$, C. Bordeanu$^1$\footnote{On leave from Horia Hulubei-National
 Institute for Physics and Nuclear Engineering, Bucharest-M\u{a}gurele,
 Romania.}, 
G. Goldring$^1$, M. Hass$^1$,  
L. Weissman$^2$, V.N. Fedoseyev$^3$, 
\\ U. K\"{o}ster$^3$, Y. Nir-El$^4$, G. Haquin$^4$,
 H.W. G\"{a}ggeler$^5$, R. Weinreich$^5$ and the ISOLDE collaboration }
\affiliation{
{ 1.~~Department of Particle Physics, Weizmann Institute of Science, 
Rehovot, Israel}\\
{2.~~NSCL, Michigan State University, East Lansing, USA}\\
{ 3.~~ISOLDE, CERN, Geneva, Switzerland}\\
{ 4.~~Soreq Research Centre, Yavne, Israel}\\
{ 5.~~{ Paul Scherrer Institute,  Villigen, Switzerland}}\\}
\date{\today}

\begin{abstract}
The $^7$Be(p,$\gamma)^8$B reaction plays a central role in the evaluation of
 solar neutrino fluxes. We report on a new precision measurement of the
 cross section of this reaction, following our previous experiment with 
an implanted $^7$Be target, a raster-scanned beam and the elimination of 
backscattering losses. The new measurement incorporates a higher activity $^7$Be
target and a  number of improvements in design and procedure. The cross 
section at a selected energy of  $E_{\rm c.m.}$ = 850 keV
was measured several times under varying experimental conditions, 
yielding a value of $S_{17}(E_{\rm c.m.}$~=~850 keV)~=~24.0~$\pm$ 0.5 eV~b,
to serve as a benchmark. 
Measurements were carried out at lower energies as well. Due to the precise
 knowledge of the implanted $^7$Be density profile it was possible to 
reconstitute both the off- and on- resonance parts of the cross section and
 to obtain from the entire set of
 measurements an extrapolated value of $S_{17}(0)$ =21.2 $\pm$ 0.7 eV~b, using
the cluster model of Descouvemont and Baye.

\end{abstract}
\pacs{PACS 26.20.+f, 26.65.+t,25.40.Lw}
\maketitle
\section{Introduction}
The study of fusion reactions in the sun, relevant to the observed 
solar neutrino shortfall, has been the subject of intensive  research, 
reviewed in references \cite{Adel,Bahc1}. Recently, this subject has acquired 
additional significance  with the new results of the Super-Kamiokande \cite{SK}
and  SNO \cite{sno}  experiments, demonstrating the existence 
of neutrino oscillations. The  $^7$Be(p,$\gamma$)$^8$B
reaction and the accurate determination of the astrophysical $S_{17}$(0)
factor is of great importance to the solar neutrino issue  and to other related
astrophysical studies \cite{Fiorentini,Barger,Lopez} since  $^8$B is
the  source of the  high-energy neutrinos from the sun that are detected
in the SNO, Kamiokande and Homestake \cite{davis} experiments.

The direct capture cross sections are measured at high energies compared to 
the solar energies (20 keV) and extrapolated to `zero' energy using an 
 energy dependent parametrization. The quantity  used  for the 
extrapolation is the astrophysical S-factor $S(E)$ which varies slowly with 
energy compared to the cross section $\sigma(E)$. $S(E)$ is related to 
$\sigma$(E) by the relation:
\begin{equation} 
S(E) = E\sigma(E)\exp[2\pi Z_1Z_2e^2/\hbar v]
\label{s17eq}
\end{equation} where $Z_1, Z_2$ are
the atomic numbers and $v$  the relative velocity.

The most widely used method of obtaining the cross section for the
$^7$Be(p,$\gamma$)$^8$B
reaction is  by direct measurement of capture of  protons on a $^7$Be target
 and the detection of the $\beta$-delayed $\alpha$ particles from the 
decay of $^8$B (see \cite{Adel} and Refs. therein). 
A direct measurement of the cross section has also been carried out with 
the kinematically inversed reaction $^1$H($^7$Be,$\gamma$)$^8$B \cite{terrasi}, 
albeit with limited statistical accuracy. 
There are also  on record  various non direct measurements of $S_{17}$ which
can be categorized as: (a) Coulomb break up of $^8$B in the time dependent 
electromagnetic field of a high $Z$ target \cite{kikuchi,iwasa,david} and
 (b) Peripheral reactions
which are amenable to the ANC ( asymptotic normalization coefficient)
 treatment \cite{tribble,azhari,trache,brown}.  
These methods, as stated in the publications above and also in a recent summary
article of indirect methods \cite{moto},  are still subject to  uncertainties 
related to the model-dependence of the extracted $S$-factor values.

In previous publications \cite {Weiss,Hass} we have demonstrated a new
method for measuring  the cross section of the $^7$Be(p,$\gamma$)$^8$B
reaction by overcoming  several of the recognized potential
 systematic errors in earlier measurements (see, e.g., Ref. \cite{Adel}).
Our method involves a small diameter implanted  $^7$Be target
from ISOLDE (CERN) and   a raster-scanned
 beam over an area larger than the target spot, avoiding  the  
difficulties  encountered with targets  of non-uniform  areal distribution.
The implanted target also eliminates the backscattering loss of $^8$B.
 Several   experiments \cite{Hammach,Jung,Stried}  have recently been
 published, quoting  
$S_{17}(0)$ values of (3-10)\% accuracy, two of those \cite{Jung,Stried} using
methods similar to Ref. \cite{Hass}. However, there still exist large,
 up to 20\% 
discrepancies among experimental results as well as the 
extracted $S_{17}(0)$ values of these measurements. 
The present work
has been undertaken in order to address these discrepancies and to
provide a new, firm input for the determination of this cross section  by
 exploiting fully  the  advantages of the implanted target: full knowledge
of the target composition and the $^7$Be profile, target robustness and the
ability to produce a secondary target of reduced activity to improve the
conditions for the $\gamma$ calibration of the target. Another feature of the
present work is a thin $\alpha$ detector and  relatively small solid angles,
 providing clean $\alpha$ spectra.

 A brief account of this work has been published elsewhere
\cite{lagy}.
Here we present the full  details of the experiment and analysis and a  
comparison to other  recent results.

\section{Experiment and Procedure}
The general scheme  of the experiment follows that of our previous 
publications \cite{Weiss,Hass}. We repeat here for convenience some sections
of Ref.  \cite{Hass} with suitable changes and additions.  
The main feature of  this measurement is  the use of a small size  target, 
implanted into a low-$Z$  material and a uniformly scanned particle beam larger
than  the target.  In general, and as described   in Ref. \cite{Weiss}, the 
reaction yield is given by 
\begin{equation}
 Y = \sigma \int  {dn_{\rm b}\over dS} {dn_{\rm t} \over dS} dS
\label{eq1}
\end{equation}
where $ n_b$, $ n_t$ are the total numbers of beam and target  particles,
respectively, and  $ dn_b/dS$, $ dn_t/dS$ are areal densities.

When the target is known  to be uniform  and the beam  is smaller  than 
the target, Eqn. (\ref{eq1}) can be simplified to the familiar relation;
\[
Y = \sigma   { dn_t \over dS} \int {dn_b\over dS} dS =   \sigma {dn_t
 \over dS} n_b
\nonumber
\]
 
In such a case, the evaluation of the  cross section depends only on the 
total beam flux. 
However, for targets of non-uniform areal distribution, e.g  radiochemically
prepared $^7$Be targets \cite{weissn},
the  full relation (\ref{eq1}) has to be used in the 
evaluation. The inherent  uncertainties  in the distributions $ dn_{\rm b}/dS$
and $ dn_{\rm t}/dS$ may  lead to  considerable uncertainties  in the 
value of the  integral  and hence  in the  deduced cross section. We have
addressed this problem  by reversing  the  arrangement: we  use a homogeneous
beam - produced by raster scanning - impinging on a target smaller  than the 
beam. The relation (\ref{eq1}) then reduces to:
\begin{equation}
 Y = \sigma {dn_b \over dS} n_t
\end{equation}
requiring  only a determination of the total  number of target nuclei and of
 the  areal density  of the beam. 

\begin{figure}\begin{center}
\includegraphics[scale =.6 ]{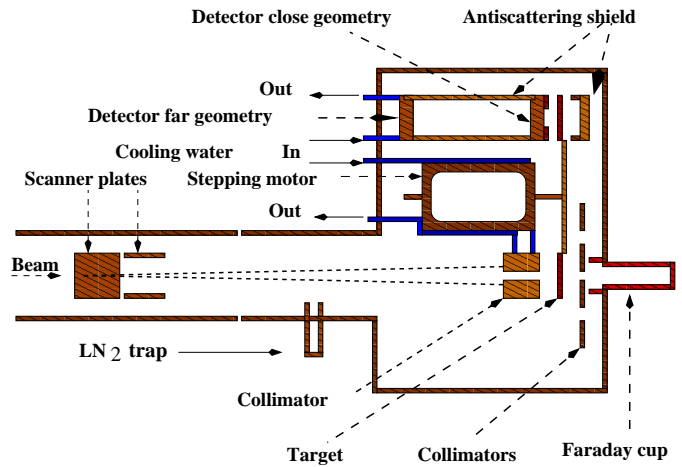}
\caption{ A schematic view of the  experimental setup.} 
\label{setup}
\end{center}
\end{figure}

The general scheme of the  experiment is  shown in Fig. \ref{setup}.  
A proton  beam from the  Weizmann Institute 3 MV Van de Graaff accelerator is 
raster scanned over a rectangle  of 4.5 mm $\times$ 3.5 mm.  The purpose of the
 scan  is to obtain a beam of uniform areal  density, as demonstrated in detail
in Refs. \cite{Weiss, Weiss1}. The scanned proton beam is collimated by a 3 mm 
diameter hole and impinges on the $^7$Be target of 2 mm diameter. A liquid
nitrogen  cooled cryofinger  is placed  close to the  target area to protect 
the target surface from contamination.  
A vacuum of $\approx$~6$\times$10$^{-7}$ mbar was maintained in the chamber.
The target spot is  aligned with a 
set of interchangeable collimators downstream from the target. The target is
mounted on an arm which is periodically rotated by a micro stepping motor 
out of the beam and in front of a silicon surface barrier detector,
registering the delayed $\alpha$'s  following  the $\beta$ decay of $^8$B.  
In the present  experiment a 150 mm$^2$, 25 $\mu$m silicon surface barrier
 detector was 
used which provided a sufficiently thick depletion layer to stop the $\alpha$
particles from  the reaction but minimized the interaction with 
$\gamma$ rays from the $^7$Be activity. The detector
was mounted at distances of 7-10 mm from the target.  

The time sequence of the entire cycle is: 1.5 s beam-on-target; 
100 ms rotation;
1.5 s target in the counting position; 100 ms rotation back to the beam 
position. The detection efficiency \cite{Filip} resulting from the known life 
time of $^8$B and this cycle is 
$\eta_{\rm cycle}$~=~0.390 $\pm$ 0.001. In the  counting position a signal from 
the motor control  unit
enables an  ADC for $\alpha$ counting and the gated scaler  for Faraday-cup 
beam monitoring. In the beam-on-target position a second ADC  is enabled for 
background counting.

In terms of the experimental parameters, the cross section of the reaction
$^7$Be(p,$\gamma$)$^8$B $\rightarrow$ $^8$Be $\rightarrow$ 2$\alpha$ 
 can be written as:
\begin{equation}
\sigma( E_{c.m.}) = {N_\alpha\over n_t} \left( {A \over N _p}
\right)
{1 \over \eta_{\rm  Be}\times\eta_{ \rm cycle}}
\label{nalpha}
\end{equation}
where $N_\alpha$ is the number of measured $\alpha$ particles,
  $N_p/A$
is the integrated current density through a collimator hole of area $A$ and
 $\it \eta_{\rm Be}$ is the geometrical detection efficiency
of the reaction products (see \ref{solid}), which
is twice the detection efficiency of $\alpha$'s.

\begin{figure}\begin{center}
\includegraphics[scale =.6 ]{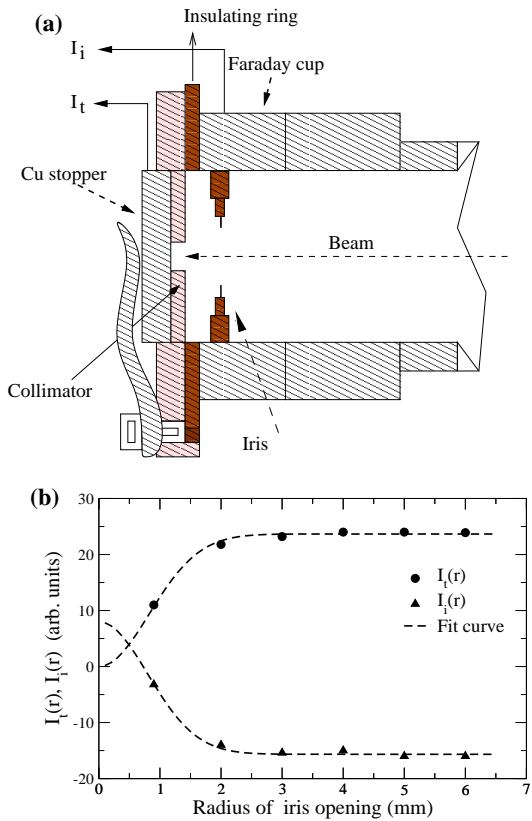}
\caption{(a): A schematic view of the  implantation chamber. Note the 
collimator of 2 mm diameter in the beam path that defines the dimensions of the 
$^7$Be spot. 
(b):  The measured beam profiles as function of the iris opening.}
\label{implant}
\end{center}
\end{figure} 

\subsection{The implanted target} \label{implant_t}

The $^7$Be target was prepared at ISOLDE (CERN) in a  manner similar to 
that  described in Ref. \cite{Hass} by direct implantation 
of $^7$Be at 60 keV in a copper disc of 12 mm diameter and 1.5 mm thickness.
The main novel feature was the primary source of $^7$Be for  ISOLDE:
a graphite target from the Paul Scherrer Institute (PSI) \cite{PSI},
used  routinely  at PSI for the production of $\pi$-mesons.
A large number of spallation  products is accumulated in the target, including
$^7$Be. A fraction of this graphite target was placed inside an ISOLDE target 
container and brought to ISOLDE for an off-beam implantation using a resonance
ionization  laser ion source. The laser ionizes $^7$Be selectively but $^7$Li,
present copiously in the target, is  also selected as part of the  mass 
selected $A = 7$ beam; $^7$Li is surface ionized in the  hot tungsten 
 cavity  due to its low ionization potential. The 
ratio of  $^7$Be to $^7$Li can be measured by switching off the laser, leaving 
only the $^7$Li ions. The initial $^7$Li current was quite high (of the order 
of 100-200 nA) but after heating the source for a few hours, the $^7$Li current
dropped considerably and an almost pure beam of $^7$Be (40-90 nA) was 
obtained. The average $^7{\rm Li}/^7{\rm Be}$ ratio during implantation was 
$\approx$ 0.08. Subsequent, more precise 
measurements of the $^7$Li content yielded a value of 0.11 for this ratio at
t=0, the end of implantation on December 6, 2001.

The implantation was carried out in a way that provided full control of 
implanted $^7$Be and $^7$Li as we all a determination of the areal density 
of the implanted ions. The total number of implanted ions was determined by 
recording the integrated charge of beam-on-target, and the ratio $^7$Li/$^7$Be
was determined by repeated measurements of the ion current with the laser 
`on' and `off'.

Credible current measurements depend on Faraday cups or other means of 
secondary-electron suppression, all requiring several centimeters of space 
between the defining collimator and the target. The postulated sharp and 
accurate definition of the target spot dictates, on the other hand, close
proximity of the defining collimator to the target.  This problem was solved 
in an implantation chamber shown schematically in Fig. \ref{implant}A. The
chamber contains two equipotential regions: 
region 1, the first along the beam, has an iris diaphragm with an opening 
radius $r$  of 0.9-6.0 mm and attached 
to it a Faraday cup.  Region 2 contains the target button: a copper disc 12 mm
 in diameter and 1.5 mm thick, pressed against a thin steel plate with a 2 mm 
diameter hole in the center, the defining collimator.  This assembly is in turn 
pressed against the frame of the iris diaphragm with an intervening
thin insulating ring.

\begin{figure}\begin{center}
\includegraphics[scale =.6 ]{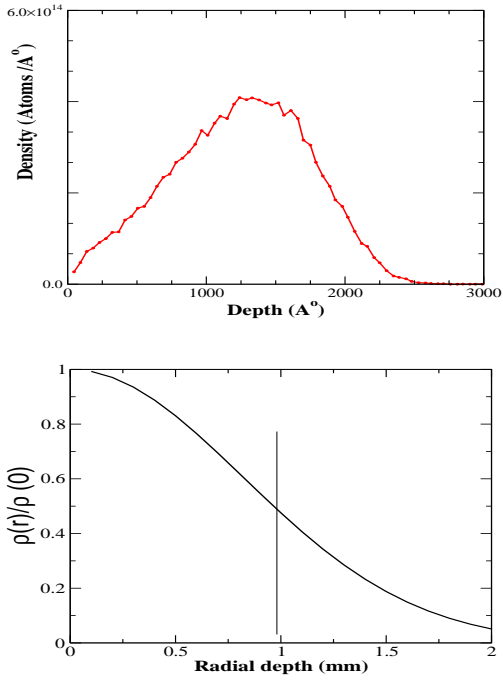}
\caption{Top: The simulated depth distribution of 60 keV $^7$Be in copper.
 Bottom: The areal density distribution  of $^7$Be beam used for implantation at
ISOLDE, this is also the $^7$Be  density distribution  in the target. 
The vertical line  indicates the edge of the implanted target.
(see \ref{implant_t} for details).} 
\label{tdist}
\end{center}
\end{figure} 
Secondary electrons from the  ions hitting the iris at opening $r$, $n_i$(r),
are collected in
the Faraday cup at the same potential, and such ions are therefore
recorded faithfully.  Secondary electrons from the $n_t$(r) ions hitting the 
target assembly are also  collected in region 1 and
they enhance the current readings from region 2 and diminish by the
same amount the current readings from region 1.  If we denote $I_i$(r) and
$I_t$(r) as the integrated currents from the two regions, and $\alpha$ as the
secondary emission coefficient, we have:
$$  I_i(r)=n_i(r)-\alpha n_t(r),~~~ I_t(r)=(1+\alpha) n_t(r)$$
$$ I_i(r)+I_t(r)=n_i(r)+n_t(r)=n_{\small T}$$ $n_T$ being the total number
 of ions in the beam.
Fig. \ref{implant}B shows the measured $I_i$(r) and $I_t$(r). The dashed lines
are fits to the points  with functions:
$I_t(r)~=~I_t(\infty)(1~-~\exp(-{r^2\over a^2})),~~  I_i(r)~=~n_T-I_t(r)$ with
 $n_T$=7.5, $a^2$=1.3 mm$^2$.
The areal density of the $^7$Be beam, $\rho(r)$ and ipso facto
 of the target can be  computed as
 $\rho(r) = \rho(0)\exp(-{r^2\over a^2})$,
which is shown in Fig. \ref{tdist} (bottom).

The recording of implanted ions is carried out in two steps:  (a) from the
profile measurement one gets for $r = 1$ mm:
$$R={I_t(1)\over
I_t(\infty)}={(1+\alpha) n_t(1)\over (1+\alpha) n_t(\infty)}={n_t(1)\over
 n_T}$$
$R$ was measured to be  $R = 0.63 \pm  0.1$.
(b) during the implantation the sum of $I_i(r)$ and $I_t(r)$ was measured with
a large opening  and integrated, yielding the  number of $^7$Be ions collected 
in a  disc of 1 mm radius  at the end of implantation (t = 0):
$$
n_t(1) = n_T\times{I_t(1) \over I_t(\infty)}  = (1.18\pm0.12) \times 10^{16} 
$$
The precise determination  of the  number of $^7$Be using $\gamma$ counting
(see \ref{tcalib}) yielded the value (1.168 $ \pm $ 0.008)$\times$10$^{16}$ for
this quantity.

\begin{figure}\begin{center}
\includegraphics[scale =.4]{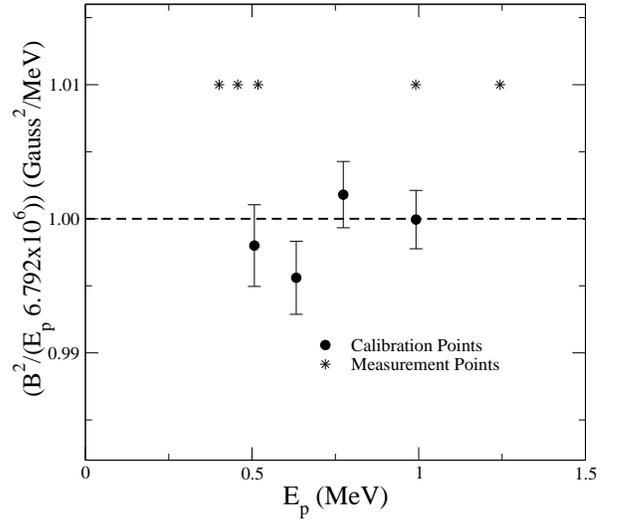}
\caption{Energy calibration of the Van de Graaff accelerator. The circles 
represent the  calibration points and the  `$*$' represent the energies at
which $^7$Be(p,$\gamma$)$^8$B measurements were carried out. The constant
 (6.79 $\pm$ 0.02)$\times$10$^6$ Gauss$^2$/MeV used to 
scale the Y-axis is obtained from the fit  as described in the text 
(section \ref{calib}).}
\label{vancal}
\end{center}
\end{figure} 
The implanted $^7$Be target   has a number of  important advantages:
\begin{enumerate}
\item{  The 
implantation profile is known from simulation (TRIM) and from a 
direct Secondary 
Ion Mass Spectrometry (SIMS) measurement for $^9$Be implanted in Cu \cite{Hass}.
 Fig. \ref{tdist} (top) shows the density  distribution of 60 keV $^7$Be's
implanted in copper. One parameter
of this simulation - the depth of the centroid of the distribution - has  now
also been  confirmed in the present measurement (see \ref{thist}).
The knowledge of the composition and  the $^7$Be density profile
is important for   evaluating the   backscattering loss of $^8$B's
from the  target. We computed the $^8$B backscattering loss to be 0.2\%, 
small enough to be ignored. The areal 
distribution of $^7$Be in the target is also known (Fig. \ref{tdist} (bottom)).}
\item{ The elemental composition 
of the  target is known precisely; the target consists of copper, $^7$Be
 and $^7$Li.  We had immediately after implantation  (t=0)
1.17$\times$10$^{16}$ $^7$Be atoms and 10$^{15}$ $^7$Li atoms in a 
cylindrical
volume of copper, 2 mm in diameter and 2500 $\AA$ deep, containing a total of 
6.7$\times$10$^{16}$ Cu atoms. The majority of the  $^7$Li atoms during the 
time of the experiment were $\beta$-decay  daughters  of $^7$Be and they have
the  same density profile.} 
\item{The target is robust. We have  direct evidence that
both the $^7$Be and $^7$Li atoms remained stable in the Cu matrix  throughout
 the duration of the experiment (see \ref{thist}), with the exception of a 
singular event - `the thermal episode' to be described later.}

\item{
  The target is calibrated for $^7$Be content by monitoring
 the  $\gamma$ rays following the  $^7$Be$~\rightarrow$~$^7$Li $\beta$ decay. 
The $\gamma$ measurements carried out in standard counting arrangements
require the $\gamma$ activity of the  sample to be  below a  
limit much smaller than the actual  activity of our target. With the 
implantation technique it was possible to produce for the purpose of 
calibration a secondary target about
300 times weaker than  the primary target, identical to it in all other aspects
 (cf. \ref{tcalib}). }
\end{enumerate}
Based on the  geometrical  parameters of the target 
(Fig. \ref{tdist} (bottom)) one can  
estimate the sensitivity  of the cross section measurements  to  beam 
inhomogeneity. Assuming  a target  distribution $e^{-{r^2\over a^2}}$ and  a
 beam
distribution $e^{-{r^2\over b^2}}$, one can compute the ratio $X_R$ of the true 
cross section as computed from Eqn. (\ref{eq1}) to the value obtained with
the procedure of Eqn. (\ref{nalpha}). For example, with $a^2 = 1.3~{\rm mm}^2$
and $b^2 > 0.5$ one gets to  a good approximation: $X_R = 1 + {0.06 \over b^2}$.
This implies a rather low sensitivity  to beam inhomogeneity. In an extreme case
of $b^2 = a^2 = 1.3~{\rm mm}^2$ one gets: $X_R = 1.05$.  One can  also well 
account in this way for the seemingly  low reduction in the value of 
${N_\alpha \over CI}$ when the beam scan is switched  off (Fig. \ref{scan}). 

\subsection{Proton energy calibration} \label{calib}

The proton energy of the  Van de Graaff accelerator was calibrated  with the 
$^{27}$Al(p,$\gamma$)$^{28}$Si resonances  at energies of 991.2, 773.7, 632.6
and 504.9 keV.
The calibration curve for the accelerator is shown in Fig. \ref{vancal}. The
constant $B^2/E_p$=(6.79 $\pm$ 0.02) $\times 10^6$ Gauss$^2$/MeV, 
where $B$ is the field of 
the analyzing magnet, is seen to  be in excellent
agreement with all the measured  points and has been used to establish
intermediate points of $E_{\rm lab}$ as well as  lower values, down to 
$E_{\rm lab}$ = 430 keV and one higher point at $E_{\rm lab}$ = 1244 keV.

\subsection{Beam uniformity}
\begin{figure}\begin{center}
\includegraphics[scale =.6]{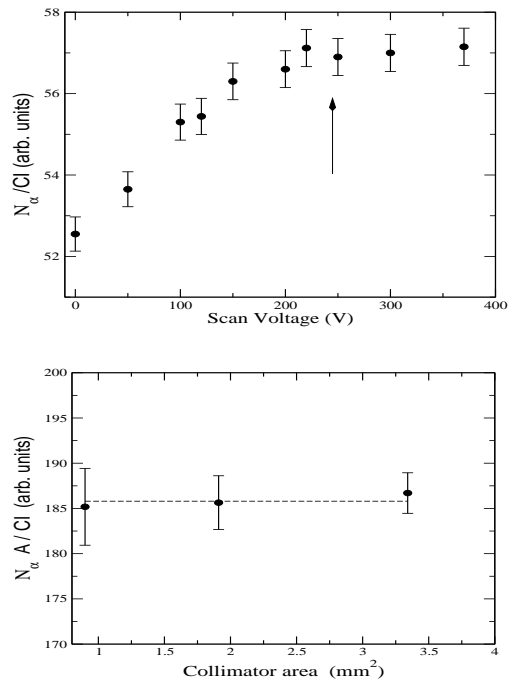}
\caption{Top: $\alpha$ yield from the $^7$Li(d,p)$^8$Li reaction at 
$E_d$~=~770 keV, normalized to
the integrated beam  current as a function of the scan voltage. The scan
voltage used in the experiment, scaled to the corresponding beam energy, 
 is indicated by the  marker.
Bottom: $\alpha$ yield from the $^7$Li(d,p)$^8$Li reaction for various 
collimators normalized to the
integrated beam  current and multiplied by the  area of the collimator.}
\label{scan}
\end{center}
\end{figure}

 The scanned 
beam density $ dn_b/dS$, typically of about 0.3-0.5 $\mu$A/$mm^2$ 
was measured by integrating the  beam in an electron  suppressed
Faraday cup after passing through  a 2 mm diameter aperture. The  current
 was digitized and recorded  in a 
gated scaler. Beam integration with and without suppression yielded results 
similar to within a fraction of a percent. 
The beam uniformity was checked as in Ref. \cite{Weiss}, by measuring  the 
$\alpha$ yield ($N_\alpha$) from the $^7$Li(d,p)$^8$Li reaction versus 
integrated current (CI) for  three downstream collimators with nominal diameters
2, 1.5 and 1 mm. Fig. \ref{scan} shows the measured
 ${N_\alpha \over CI}\times A$. 
The constancy of  this quantity for the different  collimators attests  the 
constancy of the average  number of beam particles (deuteron)  in the 
circular area of the beam (see also \ref{solid}). To obtain the optimum scan
voltage, the $\alpha$ yield from the $^7$Li(d,p)$^8$Li 
reaction at $E_d$ = 770 keV was measured as a function of scan voltage 
(Fig. \ref {scan} (top)). 
This procedure was
repeated at a lower energy (470 keV) also to  obtain the corresponding scan
 voltage.
A correction was  applied to  the measured beam density for  the finite
distance between the target and  the beam collimator and the diverging beam  
from the scanner plates. The 
beam collimator was positioned 10 mm  behind the target and the scanning plates
to target distance was 140 cm, yielding a  correction factor of 1.014.

\subsection{Target calibration} {\label{tcalib}

The $^7$Be content  was determined as in Ref. \cite{Hass}  by a measurement of
the $\gamma$ activity of the target   using known values of the  branching 
ratio of the $^7$Be $\beta$ decay to the first excited state of $^7$Li and the 
$^7$Be half life.
The $\gamma$ activity  of the $^7$Be target  was too  intense to be assayed
in a standard  $\gamma$ calibration setup due to the problems associated with
large dead times in the $\gamma$ counting.  As stated above,
a weak  target  was prepared for $\gamma$ calibration. An
accurate measurement of the relative intensities of the $^7$Li  478 keV 
$\gamma$ transition for the two targets was  carried out at the low background 
$\gamma$ counting laboratory of  NRC-Soreq by placing both at the same
distance  from a Ge counter, yielding  a ratio of 317.8 $\pm$ 0.8. This
 ratio was remeasured  several times with consistent results. The absolute 
intensity of the weak target was measured at the NRC-Soreq laboratory and also
at Texas A$\&$M University.  Both measurements followed calibration procedures
involving up to thirteen high-precision standard sources of ten radio-nuclides.
The Texas A$\&$M measurement also incorporated a precise $^{60}$Co source
(0.1\% error on its absolute activity) and
 Monte Carlo calculations which agreed 
to within  0.2\% with all measured data from 50 - 1500 keV  for  interpolating
between measurements \cite{hardy}. 
 The two measurements yield a $^7$Be content for the weak target of 
 (2.667 $\pm$ 0.018) $\times$ 10$^{13}$ and (2.650 
$\pm$ 0.018) $\times$ 10$^{13}$,
 respectively. The number of $^7$Be  nuclei in the target after implantation
(t=0) was determined to be 
$n_t$~=~(1.168~$\pm$~0.008)~$\times$~10$^{16}$. In this evaluation, the 
branching ratio for gamma emission in the decay of $^7$Be was taken to be
(10.52 $\pm$ 0.06)\% \cite{ndata} and the half-life was taken as 
 53.29 $\pm$ 0.07 days \cite{tuli}. The error on the branching ratio contributes
the major fraction of the  error  on the total number of $^7$Be nuclei.
\subsection{Cross section measurement with a thin target of finite width}
For a target of finite width, one has to  account for the fact that the beam 
energy is spread over a finite range. For a thin target this  can be 
accomplished  by
characterizing  the target density distribution $\rho$(x) by  moments:
\[
M_0 = \int_{x_0}^\infty \rho (x) dx = n_t,~  ~
 \nonumber
\]
\[
M_1 = {1 \over n_t} \int _{x_0}^\infty x \rho(x)dx = \overline x,
\]
where $\overline x$ is the~ centroid~ and
\[
M_2 = {1 \over n_t} \int _{x_0}^\infty (x- \overline x)^2 \rho(x) dx
\]
 The proton energy is related to the depth by:  

\[ E = E_0-\kappa x
\]
where  $\kappa = dE/ dx$, which  is 
 assumed~ to be constant~ for~ a~ thin~ target.

The cross section $\sigma$ close to some point $x^*$ can be  expanded in a 
power series
\begin{widetext}
\[
\sigma[E(x)] = \sigma[E(x^*)] - \left[E(x) - E(x^*)\right] \kappa\sigma^\prime
 [E(x^*)] +
 {1\over2}[E(x)-E(x^*)]^2\kappa^2
\sigma^{\prime \prime} [E(x^*)]
\nonumber
\]
where $\sigma^\prime(E)$ = ${d\sigma \over dE}$ and $\sigma^{\prime \prime}$=
${d^2\sigma\over dE^2}$.
The energy averaged cross section $<\sigma>$ is given by
\[
<\sigma>n_t=\int _{x_0}^\infty \sigma(x) \rho(x)dx 
 = \sigma[E(x^*)]n_t-\kappa 
\sigma^\prime[E(x^*)] \int_{x_0}^\infty[x- x^*]\rho(x)dx+  
{1\over 2}\kappa^2\sigma^{\prime \prime}[E(x^*)]
\int_{x_0}^\infty[x-  x^*]^2 \rho(x)dx 
\]
\end{widetext}
If $x^*$ = $\overline x$ and  $\overline E$ = $E(\overline x)$, 
 the second term vanishes and
\begin{equation}
<\sigma> =  \sigma(\overline E)\{1+{1\over 2}{\kappa^2 \sigma^{\prime \prime}
(\overline E) \over \sigma(\overline E)}M_2\}
\label{newcross}
\end{equation}
According to Eqn. (\ref{s17eq}) 
\[
~~ \sigma(E)=S(E)\times{1 \over E} e^{-\sqrt{{E_G\over E}}};
\]
where $E_G = (2\pi e^2 Z_1Z_2/\hbar c)^2/2$.
Taking $S(E)$ to be  constant for small energy intervals
\begin{eqnarray}
<\sigma>& =& \sigma( \overline E)\{1+{1 \over \overline  E^2}({E_G\over 
\overline E} -7\sqrt{{E_G\over\overline  E}}+8)
\left({d \overline E_{\rm lab}\over dx}\right)^2M_2\} \nonumber \\
& = & \sigma(\overline E)\{1+\phi\}
\label{correction}
\end{eqnarray}
For the simulated distribution of Fig. \ref{tdist}, $M_1$ = 1220 $\AA$ and
$M_2 = 2.8\times 10^5 $$\AA{^2}$.  The correction term $\phi$ is given in table 
\ref{ttable} for the relevant proton energies. In the `thermal episode' (see 
section \ref{thermal})
$M_2$ was increased by a factor of 9. The appropriate values  of $\phi$
 are also given in table \ref{ttable}.

It is apparent that for practically all our measurements $<\sigma>=
\sigma(\overline E)$
is an adequate approximation. Only for the lowest energy, 
$E_{\rm c.m.}$ = 302 keV and
the broadened $M_2$ following the thermal episode  is there  a small $M_2$
correction.

We now examine the  resonance at $E_{\rm c.m.}$ = 633 keV. We write the 
resonance equation
in terms of the peak cross section $\sigma^r_{\rm max}$ and the width $\Gamma$, 
in the form:
\[
\sigma(E) = {\sigma_{\rm max}^r \over {1+{4E^2\over \Gamma^2}}}
\nonumber
\]
taking the zero of the energy scale  at the  peak of the resonance, 
\[
\sigma^{\prime \prime}(E) = {2 \over {{\Gamma^2\over 4} +E^2}}
\{{4E^2 \over {{\Gamma^2\over 4}+E^2}} -1\}\sigma(E)
\nonumber
\]
$\sigma ^{\prime \prime}$ has a sharp minimum at the maximum of the  cross
section at $E = 0$  and we get from Eqn. (\ref{newcross})
\begin{equation}
<\sigma>_{\rm max}~ =~ \sigma_{\rm max}^r\{1-{4\kappa^2\over \Gamma^2}M_2\}
\label{resnew}
\end{equation}
For $M_2$ =  2.8$\times10^5$ ${\AA}^2$, $\kappa$ = 12.3 eV/$\AA$ and 
$\Gamma$ = 34 keV, one obtains
\begin{equation}
{4\kappa^2 \over \Gamma^2}M_2 = 0.15
\label{sigresc}
\end{equation}
From Eqns. (\ref{resnew}), (\ref{sigresc}) one can derive
 $\sigma_{\rm max}^r$ 
of the resonance  from the  measured $<\sigma>_{\rm max}$.

\begin{table}
\begin{center}
\caption{The correction term to the cross section due to the second moment
of the distribution. $\phi$ is defined in Eqn. (\ref{correction}). Phase 1
and phase 2  refer to the periods before and after the thermal episode.}
\begin{tabular}{lcccc|}
\hline &  \\
  &   &   ~~~~~~~~~~~~~  $\phi$ &\\
$E_{\rm c.m.}$(keV) ~~ & ~~ (${dE_{\rm lab} \over dx}) {eV/\AA}$~~ &
~~ Phase 1~~&Phase 2 \\
\hline \hline &\\
1078 & 9.6   & -1.2$\times 10^{-5}$ & -1.1$\times 10^{-4}$ \\
850  & 10.9  & -2.3$\times 10^{-5}$ & -2.1$\times 10^{-4}$ \\
415  & 15.7  & -4.4$\times 10^{-5}$ & -4.0$\times 10^{-4}$ \\
356  & 16.0    & +2.2$\times 10^{-4}$ & +2.0$\times 10^{-3}$ \\
302  & 17.0    & +7.1$\times 10^{-4}$ & +6.4$\times 10^{-3}$ \\
\hline
\end{tabular}
\label{ttable}
\end{center}
\end{table}

\subsection{The solid angle}\label{solid}

The solid angle  subtended by the detector  at the  $^7$Be target was 
determined, as in Ref. \cite{Weiss}, with the aid of the $^7$Li(d,p)$^8$Li 
reaction on the $^7$Li accumulated in the implanted target. The measurements
were carried out in two steps:
\begin{enumerate}
\item{with  a deuteron beam of $E_{\rm lab}$ = 770 keV,  the ratio of $\alpha$
 counts
to the integrated current $({N_\alpha \over CI})_N$, was determined in the 
same geometrical  conditions as in the $^7$Be(p,$\gamma$)$^8$B measurement
(geometry $N$).}
\item{the detector was moved to a large distance from  the 
target, $ h$ (geometry F), and a collimator of radius $r$ was placed
 in front 
of it. The ratio $R = ({N_\alpha\over CI})_N/({N_\alpha \over CI})_F$ =
$\Omega(N)/\Omega(F)$ was measured. With $h~=~95.7~\pm$~0.1 mm and $r~=~
 4.99~\pm$~0.02 mm, the solid
 angle $\Omega(F)$ was  found to be: (8.520 $\pm$ 0.02)$\times$10$^{-3}$
Steradian.
and  from the  measured ratio $R$ the  solid angle $\Omega(N)$ was evaluated.}
\end{enumerate}
  
\begin{figure}\begin{center}
\includegraphics[scale =.40]{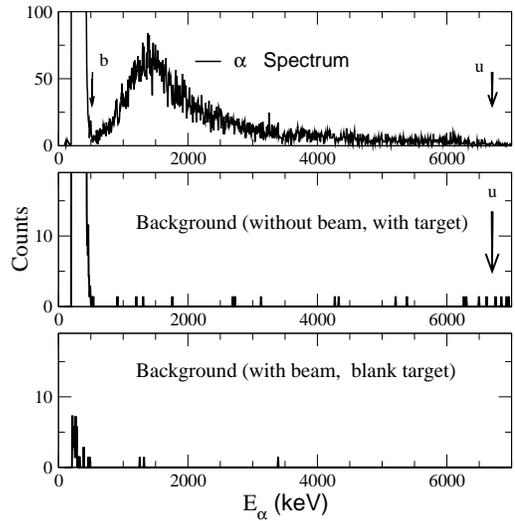}
\caption{Top: An $\alpha$ spectrum obtained at $E_{\rm lab}$ = 991 keV in 
the  geometry (1). This spectrum was collected over 
a time of $\approx$ 40 hours.  
Middle:  A background spectrum collected without beam and the target in front
of the  detector  for a time of $\approx$ 33 hours. 
The noise cut-off is at $E_{\alpha}$  $\approx$ 525 keV.
Bottom: Background spectrum collected for a  time of $\approx$ 5 
hours with the $^7$Be target replaced by a copper blank.}
\label{alphasp}
\end{center}
\end{figure} 
These steps were carried out for each individual determination of the solid 
angle in close proximity in time (not more than 16 hours between `near' 
and `far' measurements) and with  no proton bombardment in between. These
 conditions were taken as an extra precaution even though we have 
determined experimentally the overall target stability, for both $^7$Be and
 $^7$Li (see \ref{thist}). 

\begin{figure}
\begin{center}
\includegraphics[scale =.40]{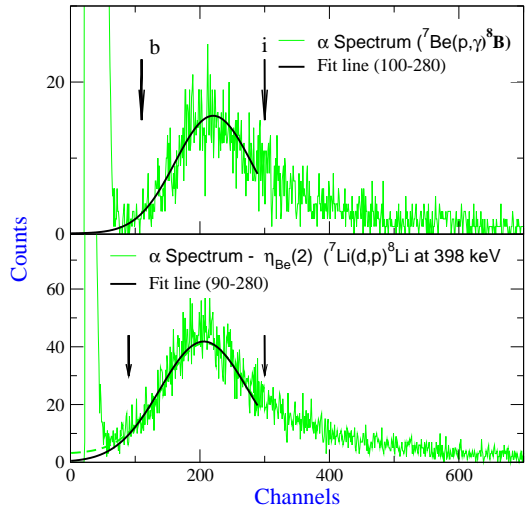}
\caption{Top: The $\alpha$ spectrum obtained from the $^7$Be(p,$\gamma$)$^8$B
reaction at 991 keV beam energy. The  line is a Gaussian fit to the 
spectrum between the arrows marked `b' and `i'.
Bottom: $\alpha$ spectrum obtained from the $^7$Li(d,p)$^8$Li reaction
for a deuteron energy of 398 keV. The Gaussian fit to the spectrum in the 
region between the arrows is  shown by the  line. In both  spectra
the extended low energy part of the fitted line agrees well with the low
energy tail of the $\alpha$ spectrum. The top panel spectrum was collected 
in the geometry (1) and the bottom one in the 
 geometry (2).}
\label{lidp470}
\end{center}
\end{figure} 
As the $\alpha$'s are emitted in pairs with $ {\bf p}(\alpha 1) +
{\bf p}(\alpha 2)$~=~0,  the efficiency for detecting  a $^8$Be 
$\alpha$ decay is twice the detection efficiency of $\alpha$'s, which
 is given by,
\begin{equation}
\eta_{\rm Be}={2\Omega_\alpha \over 4\pi}
\end{equation}

The `near geometry'  measurements were carried  out  at
four different target-detector distances of 7-10 mm. The detection efficiencies
corresponding to the  four geometries (numbered (1), (2), (3) and (4), 
respectively) used in the present experiment were: 
$\eta_{\rm Be}(1)$~=~0.1783, $\eta_{\rm Be}(2)$~=~0.2879, $\eta_{\rm Be}(3)$~
=~0.2324, and  $\eta_{\rm Be}(4)$~=~0.1752.

\begin{figure}
\begin{center}
\includegraphics[scale =.40]{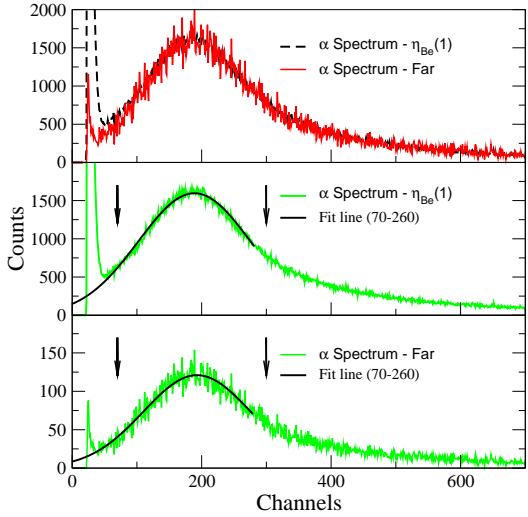}
\caption{Top: The $\alpha$ spectrum  from the $^7$Li(d,p)$^8$Li reaction
for the near geometry (1) and the corresponding far geometry scaled to 
 coincide  at the peak.
Middle and Bottom: The individual $\alpha$ spectra and the respective Gaussian
fits in the region  between the markers.}
\label{lidpspec}
\end{center}
\end{figure} 
To account for dead time, all $\alpha$ counts were referred to counts from
 a precision pulser. The system was checked by counting $\gamma$ rays from a 
standard source at a standard distance with a secondary source at varying 
distances. In general, dead times were negligible for the 
$^7$Be(p,$\gamma$)$^8$B measurements whereas 1-3 \% dead times were observed
for most of the $^7$Li(d,p)$^8$Li measurements. The highest dead times, up to 
$7\%$, were encountered in some of the near-geometry $^7$Li(d,p)$^8$Li 
measurements.

The near geometry measurements were also used to examine the overall
consistency of the  measurements. Every $\Omega(N)$ measurement
consisted of 5-7  individual high statistics measurements. These  measurements
indicated a fluctuation of $\pm$~0.8\%. Occasionally larger deviations up to
2\% were observed. These were all correlated with  indications of detuning  of
the  beam optical system and with an  abrupt change in the beam ratio $R_b$
= beam current ${{(\rm scan ~ off)}\over {(\rm scan~ on)}}$  from a 
normal 5 to about 2. The value of $R_b = 5$ is valid for both proton and 
deuteron beams, indicating that the size of  the beam spot is similar.
$R_b$ was checked routinely to monitor the beam stability. A common error of
$\pm$0.8~\%, obtained from the standard deviation of the set of 
$({N_\alpha\over CI})_N$ and $({N_\alpha\over CI})_F$ of  the $^7$Li(d,p)$^8$Li 
measurements, was applied to all individual  measurements.

\subsection{Background} \label{backg}
There are four potential sources of background: detector noise, pile up 
noise from the $^7$Be $\rightarrow$  $^7$Li$^*$ $_{\hskip 2mm}$
 $\gamma$ rays impinging on the detector,  general
background (no beam, no target) and multiply scattered protons. In our previous
experiment \cite{Hass} the detector  was found to heat up in the
 presence of the target, leading to  an increased  noise level.
In the present experiment the detector was water cooled and the  noise level
 was low and  independent of the  beam power input. The pile up
noise is a prominent  feature in the spectra, appearing as an almost 
vertical wall at the low energy end.  It is made up of high order  coincidence
events of very low  energy pulses, generated by electrons that are created
by the intense $\gamma$ radiation from the $^7$Be target.  The position of 
the endpoint of the noise wall is very sensitive to the  $\gamma$ rate
 on the detector and one observes a substantial retraction of the 
wall  with time following   the natural decay of the target, and with a 
reduced solid angle.
 The high statistics 
measurements (Fig. \ref{alphasp}) exhibit good separation between
the pile up wall and the $\alpha$ spectrum. These measurements were
analyzed in detail and  used as templates for the  low count measurements.

The general background was measured off-beam with the target in position in 
front  of the  detector. Typical spectra are shown in Fig. \ref{alphasp}.
Above the pile up wall there is a roughly even distribution of counts,
except for the high energy end where the density is appreciably  higher.
The counts above the pile up wall  are probably due to $\alpha$'s from
(n,$\alpha$) reactions in the detector or its close surroundings, and
to $\alpha$-radioactivity in the same region. The excess of counts at the 
high-energy end of the spectrum is a saturation effect due to $\alpha$'s of 
sufficiently high energy  to traverse the detector.
The  individual rates of the general background 
in  the "region of interest" range from 0.4 to 0.6 $\pm$ 0.06/hour.

\begin{figure}
\begin{center}
\includegraphics[scale =.40]{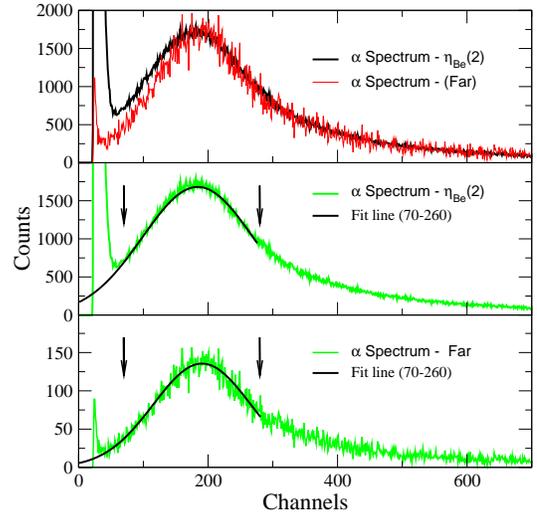}
\caption{Top: The $\alpha$ spectrum  from the $^7$Li(d,p)$^8$Li reaction
at  near geometry (2) and the corresponding far geometry scaled to  coincide 
 at the peak.
Middle and Bottom: The individual $\alpha$ spectra and the respective Gaussian
fits in the region between the markers.}
\label{lidpclf}
\end{center}
\end{figure} 
The reaction chamber was constructed with tight anti-scattering shields
around the beam and  around the  target detector assembly in the 
counting position.  To get a measure of the  scattered protons reaching the 
detector we performed a  number of  background measurements  in which the 
pile up noise was absent or reduced:

\begin{enumerate}
\item{an in-beam measurement  in the beam-on-target phase. In this case one 
expects an enhanced multiple scattering  because  the detector lacks the
 important shield
 provided   by the extended target arm,}
\item{an in-beam measurement with a reversed target,}
\item{an in-beam measurement with a blank target.}
\end{enumerate}
In the third case  the pile up noise is absent. In the first and second 
cases it  is reduced because  the 
$\gamma$ rate  at the detector is reduced due to the  larger target-detector
distance and the absorption in the copper.

In all these measurements the background level was found to be fully
consistent with the  off beam level. Furthermore, the multiple scattering
events are expected to appear at the low end of the  spectrum and to exhibit
a sharp increase with decreasing energy. No such feature was observed (Fig.
\ref{alphasp}). The small peak at the  low energy end of
the bottom  panel  of Fig. \ref{alphasp} is the  tail end  of the electronic
noise.
We therefore conclude that  in our `region of interest' there are no 
multiple scattering events.

\begin{figure}
\begin{center}
\includegraphics[scale =.40]{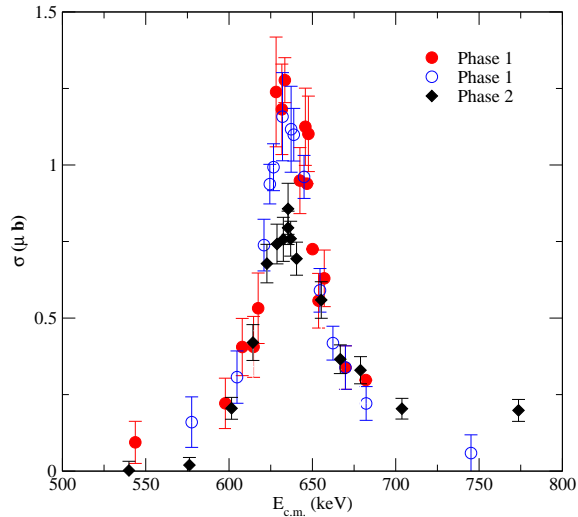}
\caption{Measurements around the  633 keV resonance in phase 1 and phase 2.}
\label{resonall}
\end{center}
\end{figure}

\subsection{Analysis of the spectra}\label{analspec}
Fig. \ref{alphasp} shows a  high statistics $\alpha$ spectrum at the 
geometry (1)
together with a background spectrum. Above the  marker 'u' the number of counts
in both spectra  is nearly the same: 22 and 17  respectively. 
The difference is negligible  compared to the  total of $\approx$ 10$^4$.
The marker 'u' was
 therefore adopted as the upper limit of the  region of interest. A  similar 
procedure was adopted for the spectra corresponding to geometries
 1, 2, 3 and 4.
At the lower end of the  spectrum
the dominant concern is the pile up wall. In a number of spectra the wall 
intercepts the $^8$Be $\alpha$ spectrum at a fairly high energy and 
it is necessary to have a reliable extension of the $\alpha$ spectrum `into 
the wall'. For a very thin target and a small detector solid angle this is a
 straightforward task because the shape of the $\alpha$ spectrum  is known
 \cite{war}.  Our target,
however, has a finite  thickness causing some  energy loss of the $\alpha$'s, 
and the solid angle in some of the measurements is quite large so that  the 
energy loss varies appreciably as a function of the  $\alpha$ emission angle.
For the evaluation of this  region of the spectrum we have adopted
a  procedure similar to  the one described in Ref. \cite{Filip}. The 
spectrum in Fig. \ref{lidp470} (top), reaches down to almost $N_\alpha$ = 0
 at the
lowest energy and we take this spectrum and some other  high quality spectra,
 rather than  the `pure'
 spectrum of Ref. \cite{war}, as  models for  the  entire set of 
measurements. A Gaussian with an area equal to  the integrated
spectrum  provides  an excellent  fit to the data from channel 280 down to 
the low end of the spectrum. The same is   true  for other high quality spectra,
e.g. as in  Fig. \ref{lidp470} (bottom), $\alpha$'s from  $^7$Li(d,p)$^8$Li
at a near geometry for a deuteron energy of 398 keV
where the $^8$Li are located closer to the surface due to the lower deuteron
momentum. We  conclude that if a marker `b' is set at an energy higher
than the pileup wall, a Gaussian fit to the region between markers `b' and `i'  
will provide an adequate extension to lower $\alpha$ energies. 

\begin{figure}\begin{center}
\includegraphics[scale =.41]{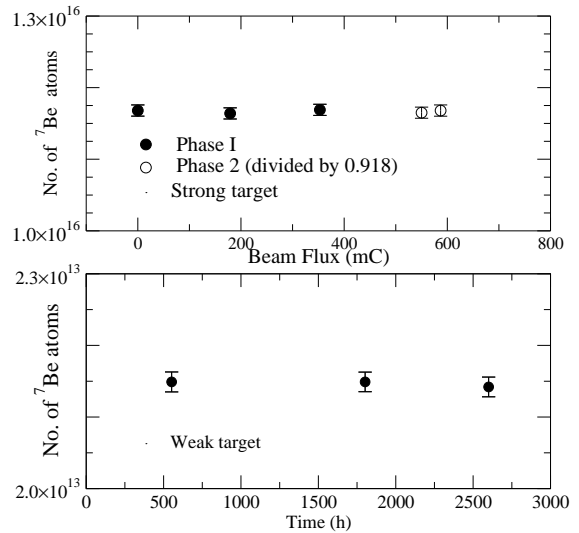}
\caption{Top: The $^7$Be target content, determined by measuring the  activity 
in a Ge detector, 
as a function of the accumulated charge on target 
 and corrected for
the decay of $^7$Be. The  open circles  are scaled by a factor of 0.918
(see text). Bottom: The number of $^7$Be atoms corrected for decay in the weak
target as function of time.}
\label{decay}
\end{center}
\end{figure}
 
The procedure for any given $^8$B spectrum is to sum the counts
 between the  `b' and  `u' markers as presented in Fig. \ref{alphasp}. 
The  excess of $\alpha$ counts below the `b' marker is then
derived from a Gaussian fit to the region: `b$\rightarrow$i' 
(see Fig. \ref{lidp470}). The  robustness
of this  procedure was verified  with the aid of some high quality $\alpha$
spectra. The spectrum in the middle panel of Fig. \ref{lidpspec}, for example,
was treated in this manner with `b' at various  positions beyond the low energy
minimum, yielding $N_\alpha$ values that are well within the statistical errors.
 The `tail corrections' for the  
$^8$B $\alpha$ spectra  range from (0.4 $\pm$ 0.2)\% to (1.0 $\pm$ 0.6)\%.

For the $^8$Li $\alpha$ spectra 
only the correction to the ratio $N_\alpha$(near)/$N_\alpha$(far) is of 
significance.  Fig. \ref{lidpspec} shows the far and near 
normalized $\alpha$  spectra for the  geometry (1).  The Gaussian
fits are  essentially identical. The  
correction  to $\eta_{\rm Be}(N)/\eta_{\rm Be}(F)$ due to tail correction
 can in this case
be  derived quite accurately, and  is found to be (0.8 $\pm$ 0.5)\%.
At the other extreme, for the closest  geometry (2), the Gaussian 
fits to 
the near and  far spectra are different (Fig. \ref{lidpclf}) and  the 
tail correction to the $\eta_{\rm Be}(N)/\eta_{\rm Be}(F)$ ratio is 
 evaluated in this case as 
(2.4 $\pm$ 0.8)\%.
The fitted line has non-zero intercepts in some cases, 
indicating   that there  is  a small fraction of the $\alpha$'s 
stopped in the  target. For the  $^8$B measurements the error  in the tail
correction is always small compared to the statistical error. For the 
$^8$Li measurements they constitute  the dominant uncertainty.
\begin{figure}
\begin{center}
\includegraphics[scale =.40]{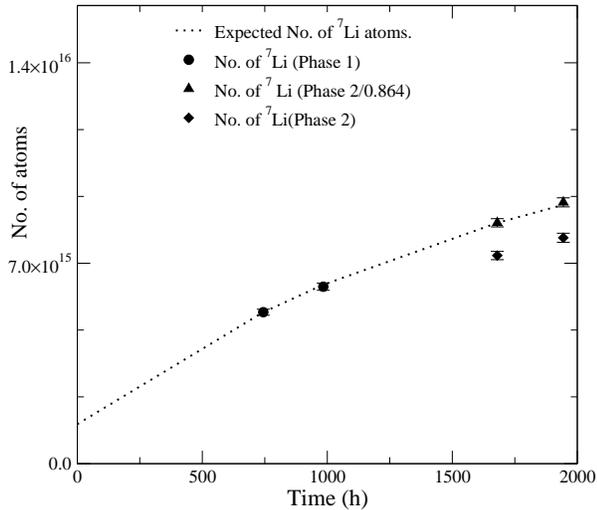}
\caption{The inventory of $^7$Li atoms in the target. The number of 
$^7$Li atoms was  measured by the $^7$Li(d,p)$^8$Li 
reaction. The circles represent the measured numbers in phase 1 and the 
triangles in phase 2,  corrected for the fraction lost in the thermal
episode.  The line through the points represents the expected $^7$Li  
accumulation  due to the  $^7$Be decay.} 
\label{lidecay}
\end{center}
\end{figure}

\subsection{The 633 keV M1 resonance}
Measurements of the resonance were carried out repeatedly to establish
and monitor the centroid ($M_1$) and the width ($M_2$) of the $^7$Be depth
distribution.  The earliest of these measurements are shown in Fig.
 \ref{resonall}.
The centroid shift of 15 keV  confirms the centroid depth of the  simulated 
distribution
at 1215 $\AA$. The value of 
$\sigma_{\rm max}^r$ can be  inferred from the measured cross section
at the peak and the value of $M_2$: 2.8 $\times$ 10$^5$ $\AA$$^2$ according to 
Eqns. (\ref{resnew}), (\ref{sigresc}) yielding  the value of
 $\sigma_{\rm max}^r$ = 1340 $\pm$ 100 nb.  
In a more elaborate evaluation presented later in section \ref{thist},
  the parameters
of the resonance were determined as: $\sigma_{\rm max}^r$~=~1250~$\pm$~100 nb
and $\Gamma$~=~35~$\pm$~3 keV.

\begin{figure}
\begin{center}
\includegraphics[scale =.37]{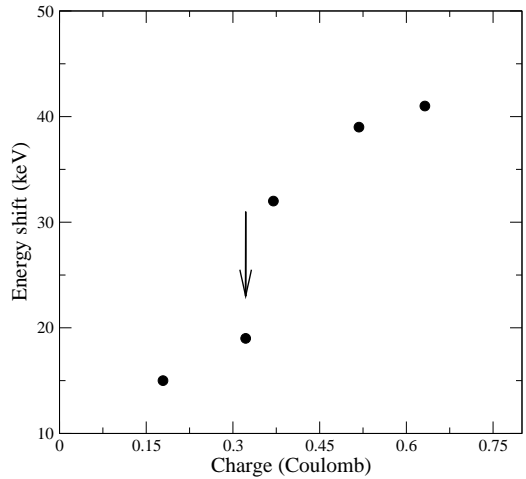}
\caption{The  energy shift of the centroid of the 633 keV resonance
 due to  carbon buildup on the target, as a 
function of the integrated charge.
 The marker indicates  the ``thermal episode'' (see text).}
\label{respeak}
\end{center}
\end{figure} 

\subsection{The thermal episode} \label{thermal}
At  350 hours into the experiment, the $^7$Be target underwent a 
`thermal episode'. A short experiment concerning the cross section of the 
reaction $^7$Be($^3$He,2$\alpha$)2p was carried out with our $^7$Be target
at the  Van de Graaff Laboratory \cite{ralph}. In this experiment the target was
not  cooled sufficiently and the $^3$He beam energy was dissipated by 
radiation.
We estimate that the target reached a temperature of at least 400$^o$C.
Following the event it was found that the $^7$Be  content of the target was
 reduced by a 
factor of 0.918 $\pm$ 0.003, the $^7$Li content  by 0.86 and the second 
moment of the  $^7$Be
distribution  was increased by a factor of 9. There was also an additional
deposit of carbon of $\approx$ 3000$\AA$.  These matters are discussed in
detail below. The accuracy and reliability of the  data were not affected by 
this, although the 
quality of the spectra at the low energy end was impaired by the extra  
carbon deposit.

The fact that after such extreme heating the target was at all usable bears 
impressive evidence to the stability of the implanted target in both the $^7$Be
and $^7$Li components. The temperature of the target was measured under
conditions of normal usage and it was established that the target was never 
 hotter than 
110$^o$C. Of the 90$^o$C increase over room temperature, roughly half is
due to the stepping motor to which the  target is thermally connected,
and the other half  - to  the proton beam.

\begin{figure}
\begin{center}
\includegraphics[scale =.40]{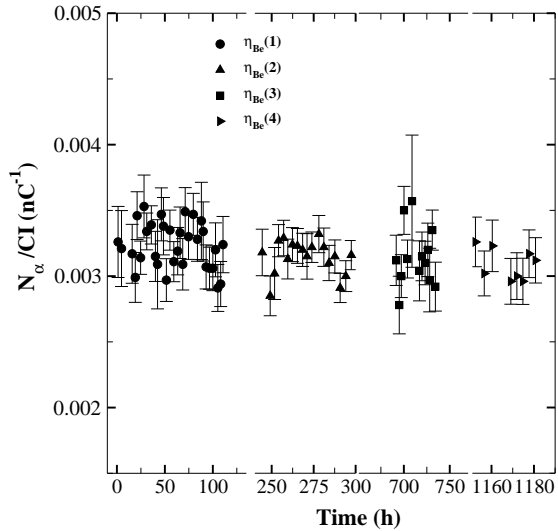}
\caption{The  ratio of $\alpha$ counts to integrated current,
 $N_\alpha$/CI, measured at the 991 keV beam energy as a function
of time and normalized to the known decay rate of $^7$Be, for various
  geometries of the $\alpha$ detector. The four symbols 
correspond to the measurements of points  2, 3, 4 and 5, respectively,
 in table \ref{stable}. 
The measurements span a period of more than 40 days.}
\label{alci}
\end{center}
\end{figure} 
\subsection{The target history}\label{thist}
Absolute $\gamma$ calibration measurements of the  weak $^7$Be target have been 
carried out a  number of times (Fig. \ref{decay}). They
clearly point to a  constant $^7$Be inventory.
There were a  number of `strong' to  `weak' comparisons, 
as well as  regular $\gamma$ monitoring measurements over an extended
period (also shown in Fig. \ref{decay}). It is apparent that the strong 
target also suffered no loss of $^7$Be (other than in the thermal episode).
The most likely cause  of $^7$Be loss is sputtering  induced by the  proton
beam. The total proton charge into the target during the entire period of the
measurement was about 1 Coulomb and we can therefore infer that the proton
induced sputtering under the condition of  our experiment is less than
1\% per Coulomb.
The $^7$Li content  can be inferred at the instance of every solid angle
measurement from the value of  ${N_\alpha \over CI}$,  the known solid angle 
and the known cross section. Fig. \ref{lidecay}  shows the measured  
numbers of Li atoms fitted to a curve $n_{\rm Li}+n_{\rm Be}(1-2^{t/T_{1/2}}$) 
where $n_{\rm Li}$ and  $n_{\rm Be}$ 
are the number of Li and Be atoms  at the time  of the target  preparation.
The good fit  attests the  stability of the $^7$Li content.

Measurements in the  resonance region  were carried out  routinely. Most were 
limited to the immediate neighborhood of the peak and were carried out with the
aim of monitoring the centroid of the $^7$Be distribution. Three measurements
 were  carried out over  a  sufficient range to provide  information also 
on the shape of the resonance. Fig. \ref{respeak}  presents the shift of the 
resonance centroid as a function of time. Both the  gradual shift and the large
increase  at the time of the thermal episode (at least most of it) are 
attributed to  the carbon build up.  The three full resonance curve 
measurements  are shown in Fig. \ref{resonall}.  The first two are seen to be 
consistent. 
A fit to a pure resonance, modified by  a target
with  the profile of Fig. \ref{tdist} with $M_2$ = 2.6$\times$10$^5$$\AA$$^2$ 
is shown in the upper panel of Fig. \ref{sfac}.
The resonance parameters extracted from this fit  are 
$\sigma_{\rm max}^r$ = 1250 $\pm$ 100 nb, $\sigma_{\rm max}$(total) = 1560 
$\pm$ 120 nb and
$\Gamma$ = 35 $\pm$ 3 keV . Previously quoted values are 
$\sigma_{\rm max}^r$ = 1180 $\pm$ 120 nb and  $\Gamma$ = 37 $\pm$ 5 keV 
\cite{Filip}.
The third resonance  curve in Fig. \ref{resonall}, taken after the 
thermal episode, is clearly lower and broader. In this  case 
$M_2$ was taken as a fit parameter and was found to be $M_2$~=~2.34$\times~
10^6$~$\AA{^2}$.
\begin{figure}
\begin{center}
\includegraphics[scale =.37]{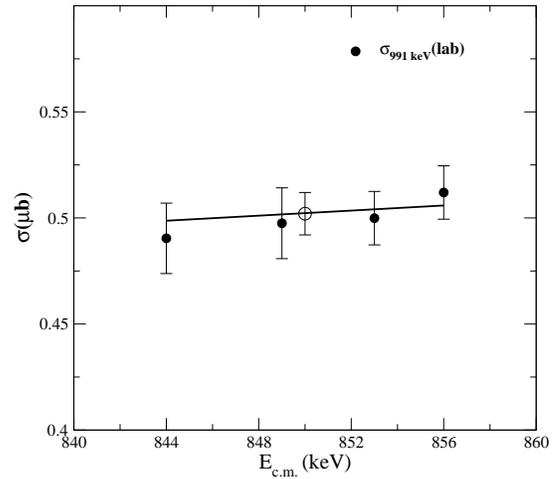}
\caption{The  cross section for $^7$Be(p,$\gamma$)$^8$B measured at 
different geometries at $E_{\rm lab}$=991 keV. The gradual energy shift is
due to 
the carbon build up on the target. The solid line is from Eqn.\ref{sigmean}
The open circle represents the average cross section 
at $E_{\rm c.m.}$=850 keV }
\label{sig990}
\end{center}
\end{figure} 

\section{The $^7$B\lowercase{e}(\lowercase{p},$\gamma$)$^8$B Measurements}

The cross section is  evaluated from  Eqn. (\ref{nalpha}).
Cross section measurements were carried out at the energies of  
$E_{\rm c.m.}$ = 1078, 856, 415, 356 and 302 keV and also around the 633 keV 
resonance. 
One of the major  objectives of this work was to carry out a measurement at one
proton energy  as accurately as possible.  This is important  in 
order to  serve as an accurate comparison  between various (future) 
measurements. Such an experimental comparison is thus free  of uncertainties
related to the extrapolation procedure of $S_{17}(0)$.
 The energy $E_{\rm lab}$=991 keV has been
chosen because this is the  energy of the $^{27}$Al(p,$\gamma$)$^{28}$Si
resonance, the  best calibration point. A number of
measurements were carried out
at this  lab energy under varying conditions of solid angle and target strength,
 with slightly varying values of $E_{\rm c.m.}$, due to carbon
buildup. The last two measurements  were performed  at the slightly higher 
energy: $E_{\rm lab}$ = 998 keV to compensate  for the carbon build up in 
the thermal episode. The  measurements were carried out
 before and after the thermal episode, and  extended altogether  over a
 period of 40 days. The four  energy points also demonstrate  the relative
 stability 
of the  measurements in the two phases.  The individual runs are shown in 
Fig. \ref{alci}  and  in Fig. \ref{sig990} and the cross section values are 
presented in  Fig. \ref{sig990} together with   a fit to the function:
\begin{equation}
 S_{17}(850 keV){1\over E_{\rm c.m.}}
e^{-\sqrt{{E_G\over E_{\rm c.m.}}}}.
\label{sigmean}
\end{equation}
  The combined value of  $S_{17}$ at this point is 
$S_{17}$(850 keV)~=~24.0 $\pm$ 0.5 eV b.
Another measurement, at $E_{\rm c.m.}$ = 1078 keV, was carried out in phase 1,
and 
three  measurements at lower energies at 415, 356 and 302 keV  in phase 2.
These  are  presented in table \ref{stable}. 

\begin{figure}\begin{center}
\includegraphics[scale =.45]{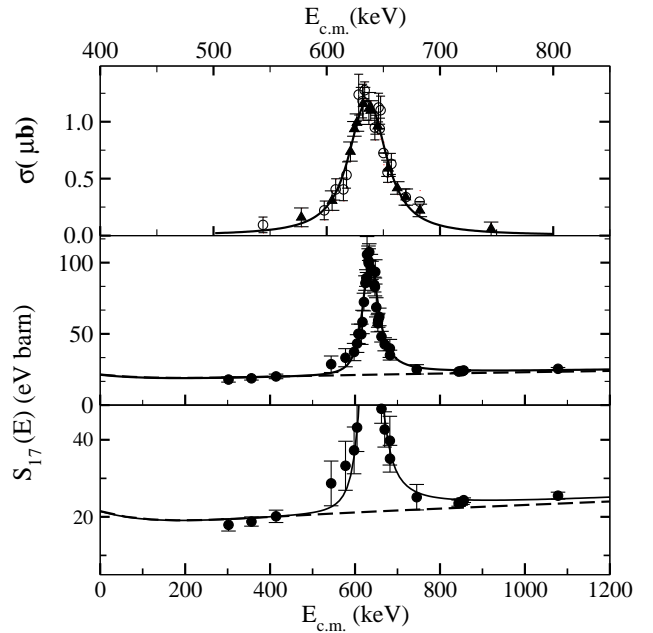}
\caption{Top:  The resonance at 633 keV. The  points are the measured
 cross sections after subtraction of the non-resonant part. The energy
axis is expanded  in comparison to the middle and lower panel.
The two symbols represent two separate measurements. The continuous 
curve represents the convolution  of a   Breit-Wigner resonance with the 
simulated $^7$Be distribution in Cu.
Middle:  The points are the measured $S_{17}(E)$.
The continuous line is  the  scaled function of Descouvemont and Baye [DB] 
  \cite{DB}  plus a 
Breit-Wigner resonance with an  energy dependent width.
 The dashed line is  the scaled DB model.
 Bottom:- An expanded view of the middle figure.}
\label{sfac}
\end{center}
\end{figure} 
Practically for all the  measurements the target can be 
considered thin even in the broadened state of phase 2, in the sense that
the integration over the density distribution amounts  to  correlating the 
cross section measurement with the  proton energy at the  centroid of the $^7$Be
distribution. Only for the 302 keV measurement does the  second term in
Eqn. (\ref{correction})  have a barely significant value of 0.6\%.
The centroid energy was determined in every instance directly for the
$^7$Be(p,$\gamma$)$^8$B resonance by measuring  the energy shift of the peak. 
This shift was then re-evaluated for the relevant energies in specific
measurements by applying the appropriate $dE/dx$ factors. For this 
one needs to know separately the amount of C and Cu in the proton  path.
We have assumed   that the  centroid of $^7$Be distribution in the
copper did not change and that  any additional energy loss  is due to  C
deposition. This assumption is well supported for  phase 1, but not so well for
phase 2. Even though  there is evidence  that a large part of the  
energy loss is  indeed in carbon, we
consider as a limit the possibility of `only Cu' for the extra energy loss.  
This yields for  the lowest proton energy, 
$E_{\rm c.m.}$~=~302 keV,  a reduction  in loss of 5 keV, a corrected energy 
$E_{\rm c.m.}$~=~307 keV and a decrease in $S_{17}$(E) of 3.7\%.
We have chosen to leave the ``only C'' value unchanged and
enlarged the error from 9\% to 10\% to cover the unlikely
eventuality. For the  higher proton energies an `only Cu' assumption does not
change  the quoted  values and errors.

The  quoted errors in table \ref{stable}   are made up of (in~\%) : 
target activity ($^7$Be content) - 0.7, background correction - 0.1 - 1.2, 
solid angle  - 0.8 - 1.4, $\alpha$ spectrum cut off 
- 0.3 - 0.8, beam collimator area - 0.8, time sequence efficiency 
- 0.2 and the uncertainty due to the proton energy 
calibration - 0.2 - 1.0. The high energy points 
were measured with statistical precision varying from 1.0 - 1.8 and the 
points below the resonance  with    (4.0 - 8.0)$\%$.

\begin{table}
\begin{center}
\caption{The measured $S_{\rm 17}(E)$ values along with  the details of the
estimated error. The  * indicates the set of measurements carried out at
$E_{\rm c.m.}$ close to  850 keV; the slightly different values of 
$E_{\rm c.m.}$ are due to gradual carbon build-up, monitored by repeated 
measurements of 
the resonance. A combined  value of  the measurements near $E_{\rm c.m.}$ = 
850 keV, using Eqn. (\ref{sigmean}) is also given. Points indicated by ** 
were measured after the thermal episode as described in the text.
Columns 3-7 represent the contributions to the  error from counting
 statistics, background, beam energy, correction for $\alpha$ loss below the
region of interest, and a set of common errors:  the
error on the number of $^7$Be atoms,  solid angle, timing efficiency, and
the area of the beam collimator.}

\begin{tabular}{cc|ccccc}
\hline
E$_{\rm c.m.}$ &~~ S$_{17}$(E)~~&~Stat. &~ B.G~ &
Energy& $\alpha$-Cutoff & Common\\
 (keV)  & (eV~b) &  & & &  &\\
\hline \hline &\\
1078    & 25.5 $\pm$ 0.8        & 0.49 & 0.03 & 0.08 & 0.20 & 0.52\\
856$^*$ & 24.3 $\pm$ 0.6        & 0.29 & 0.03 & 0.07 & 0.08 & 0.45\\
853$^*$ & 23.8 $\pm$ 0.6        & 0.24 & 0.03 & 0.07 & 0.08 & 0.48\\
849$^*$ & 23.8 $\pm$ 0.8$^{**}$ & 0.44 & 0.03 & 0.10 & 0.08 & 0.54\\
844$^*$ & 23.6 $\pm$ 0.8$^{**}$ & 0.54 & 0.03 & 0.10 & 0.14 & 0.44\\
415     & 20.2 $\pm$ 1.5$^{**}$ & 1.36 & 0.16 & 0.36 & 0.06 & 0.45\\
356     &18.8 $\pm$ 1.1$^{**}$ & 0.90 & 0.19 & 0.37 & 0.06 & 0.42\\
302     &18.1 $\pm$ 1.8$^{**}$ & 1.50 & 0.21 & 0.80 & 0.06 & 0.40\\
\hline
850     &24.0 $\pm$ 0.5~ & & & & & \\
\hline
\end{tabular}
\label{stable}
\end{center}
\end{table}

\begin{figure}
\begin{center}
\includegraphics[scale =.4 ]{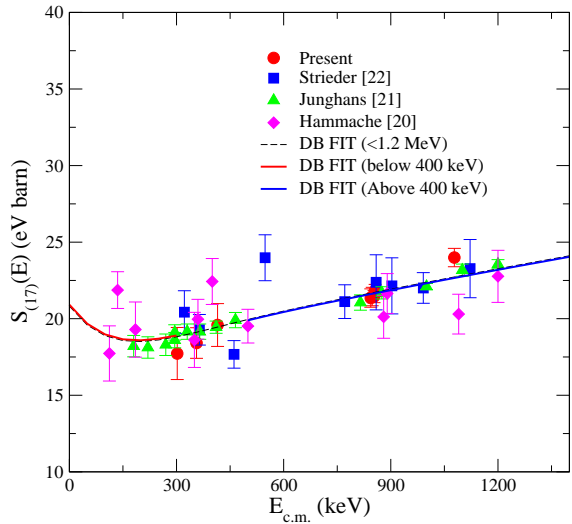}
\caption{ Non resonant part of the $S_{17}(E)$ from recent direct capture
measurements. Each set of $S_{17}(E)$ values was  fitted independently to
the DB parameterization and the individual scaling factors  were then 
renormalized to a reference value corresponding to $S_{17}(0)$ = 21.2 eV barn.
The overall consistency  of the  data upto $E_{c.m.}$ = 1.2 MeV as well as 
the agreement with the DB parameterization  is apparent.} 
\label{alldb}
\end{center}
\end{figure} 
\section{Results and Discussion}
The results of this  work are presented in table \ref{stable} in term of
$S_{17}$(E) values.
The ultimate goal of cross section measurement is to  evaluate the cross 
section at solar energies, and the  experimental determination of the 
cross section in the  region of the Gamow peak
 was and remains paramount; this goal has not been achieved yet. 
The alternative is to measure the cross section at higher energies
 and employ theoretical models for the extrapolation to solar (``zero'') 
energy. 
Several models have been proposed for  the  extrapolation of the  cross section
to  zero energy. A detailed discussion of the various models is given 
by Jennings {\it et al} \cite{jeng}, where it is shown that the various
 models coalesce at center-of-mass 
energies below $\approx$ 400 keV and that, in general, cluster models fit the 
experimental data up to $\approx$ 1.2 MeV.  The practice of employing a 
generally adopted extrapolation model is supported by the observation that the 
disagreements among experiments are mostly in factors of proportionality in the
cross section  while the measured energy dependence is largely consistent.
We have put this general observation to a  quantitative test: in
Fig. \ref{alldb} the non resonant part of the $S_{17}(E)$ values from  recent 
measurements are shown with each set of  values  fitted separately 
to the DB 
model by a  scaling  constant to yield an extrapolated values of 
 $S_{17}(0)$ = 21.2 (this value has been  chosen arbitrarily to 
match with the $S_{17}(0)$ from the present measurement).
The dashed line drawn through the data points is  a DB fit to the entire set 
with a normalized $\chi^2$ of 1.08. The energy dependence of all measurements
 is clearly similar and, in turn, fully consistent with the DB model. 
The red and blue lines 
are separate fits to the region above and below 400 keV, that agree to better
than 1 $\%$. We conclude that to the precision of the present experiments  
the DB model 
provides a  representation of the measured cross sections (applying 
the individual renormalization factors as discussed above) up to at 
least 1.2 MeV, that is as good as the ``universal'' fit of all models to 
the region below 400 keV. 

For the extraction of $S_{17}$(0)value from our measurements we have adopted 
a procedure of including all measurements, off and on the resonance,
in a fit with  the values of $\sigma_{\rm max}$ and
$\Gamma$ of the resonance and the overall normalization of the
 Descouvemont-Baye  theory as free parameters
 (Fig. \ref{sfac}).
We arrive in this way at a value of $S_{17}$(0)~=~21.2 $\pm$ 0.7 eV~b.
We quote for completeness also the value derived from the low energy 
points (below the resonance) only as: $S_{17}$(0)~=~20.8~$\pm$~1.3~eV~b. 

\begin{table}
\begin{center}
\caption{A compilation of the $S_{17}$(0) values  from recent direct capture
measurements}
\begin{tabular}{lll|}
\hline & \\
~~~Experiment~~~ &~~~ $S_{17}$(0)(eV~b)~~~ \\
\hline \hline &\\
Hammache {\it et al} \cite{Hammach} ~~~&~~~ 18.8 $\pm$ 1.7\\
Strieder {\it et al} \cite{Stried}  ~~~&~~~ 18.4 $\pm$ 1.6\\
Hass {\it et al} \cite{Hass}  ~~~&~~~ 20.3 $\pm$ 1.2\\
Junghans {\it et al} \cite{Jung}~~~&~~~~22.3 $\pm$ 0.7\\
Present  ~~~&~~~ 21.2 $\pm$ 0.7\\
\hline
\end{tabular}
\label{s17all}
\end{center}
\end{table}

Table \ref{s17all} gives the  $S_{17}$(0) 
values  of the most recent  publications of  direct capture reactions with 
a radioactive $^7$Be target. Included here are the recent 
precise  measurements (taking note inter alia of the possibility of 
backscattering loss). From table \ref{s17all} we arrive at a mean value:
$S_{17}$(0) = 21.1 $\pm$ 0.4 eV b with $\chi^2/\nu$ = 2.0 suggesting some 
discrepancy.
If we omit the  value of Ref. \cite{Jung} (which is being revised) from the
list we get a consistent  mean value: $S_{17}$(0)~=~20.5~$\pm$~0.5~eV~b with 
$\chi^2$/$\nu$ =
1.2. If we add to this in quadrature an `error in theory' of ($\pm 0.5$),
as suggested in  Ref. \cite{Jung}, we  get a consistent common value: 
$S_{17}$(0) = 20.5 $\pm$ 0.7 eV b.

When relating this value to the environment of the solar interior,
one is faced with  two inherent  uncertainties  related to the
extrapolation of the cross section to the solar energy region and 
  to the  atomic screening correction.
The $\pm$~0.5 eV b uncertainty  quoted above from Ref. \cite{Jung} is 
an attempt
to quantify the first, and the second  is believed  to be small \cite{Adel}.

The predicted $^8$B neutrino flux, $\phi$($^8$B), is directly proportional 
to $S_{17}$(0) .
If the average value  of $S_{17}$(0) quoted above is introduced into the 
Standard Solar Model \cite{bahc2}, replacing the presently   adopted value
 of $S_{17}$(0)~=~$19^{+4}_{-2}$ \cite{Adel},  the uncertainty  in
$S_{17}$(0) will   become insignificant compared to  other sources of 
error  in the evaluation of $\phi$($^8$B)  \cite{bahc3}.

We wish to thank the technical staff of ISOLDE(CERN), PSI,  and Y. Shachar and
the technical staff of the Accelerator Laboratory of the Weizmann Institute.
We acknowledge with gratitude the help of J.C. Hardy and V.E. Iacob with the
 intensity calibration of the $^7$Be target. We thank Prof. K.A. Snover for 
a fruitful exchange of information regarding the results of Ref. \cite{Jung} 
and work in progress. This work was supported in part 
by the Israel-Germany MINERVA Foundation.


\begin{thebibliography}{999}



\bibitem {Adel}{E.G.  Adelberger {\it et al}., Rev. Mod. Phys. {\bf70}, 1265
 (1998).}
\bibitem{Bahc1}{J.N. Bahcall, P.I. Krastev, and A. Yu. Smirnov., Phys. Rev. 
{\bf D 58}, 096016 (1998).}
\bibitem{SK}{S. Fukuda {\it et al}., Phys. Rev. Lett. {\bf 86}, 5656 (2001). } 
\bibitem{sno}{Q.R. Ahmad {\it et al}., Phys. Rev. Lett. {\bf 87}, 071301 (2001);
Q.R. Ahmad {\it et al}., Phys. Rev. Lett. {\bf 89}, 011301 (2002).}
\bibitem{Fiorentini}{G. Fiorentini and B. Ricci, Phys. Lett.
 {\bf B 526}, 186 (2002).}
\bibitem{Barger}{V. Barger, D. Marfatia and K. Whisnant, Phys. Rev. Lett. 
{\bf 88}, 011302 (2002).}
\bibitem{Lopez}{I.P. Lopes and J. Silk, Phys. Rev. Lett.
 {\bf 88}, 151303 (2002).}
\bibitem{davis} {R. Davis, Prog. Part. Nucl. Phys. 32, 13 (1994).}
\bibitem{terrasi} {F. Terrasi {\it et al}., Nucl. Phys. {\bf A688}, 539c 
(2001).}
\bibitem{kikuchi} {T. Kikuchi {\it et al}., The Eur. Phys. J. {\bf A 3}, 213 
(1998).}
\bibitem{iwasa}{N. Iwasa {\it et al}., Phys. Rev. Lett. {\bf 83}, 2910 (1999).}
\bibitem{david}{B. Davids {\it et al}., Phys. Rev. {\bf C 63}, 065806 (2001).}
\bibitem{moto}{T. Motobayashi, Nucl. Phys. {\bf A693}, 258 (2001).}
\bibitem{tribble} {H.M. Xu {\it et al}., Phys. Rev. Lett. {\bf 73}, 2027 
(1994).}
\bibitem{azhari} {A. Azhari {\it et al}., Phys. Rev. {\bf C 63}, 055803 (2001).}
\bibitem{trache} {L. Trache, F. Carstoiu, C.A. Gagliardi, and R.E. Tribble,
Phys. Rev. Lett. {\bf 87}, 271102 (2001).}
\bibitem{brown} {B.A. Brown, P.G. Hansen, B.M. Sherrill and J.A. Tostevin, Phys.
Rev. {\bf C 65},  061601(R) (2002).}
\bibitem{Weiss}{L. Weissman {\it et al}., Nucl. Phys. {\bf A630}, 678 (1998).}
\bibitem{Hass}{M. Hass  {\it et al}., Phys. Lett. {\bf B 462}, 237 (1999).}
\bibitem{Hammach}{F. Hammache {\it et al}., Phys. Rev. Lett. {\bf 86}, 3985 
(2001).}
\bibitem{Jung}{A.R. Junghans  {\it et al}., Phys. Rev. Lett. {\bf 88}, 041101 
(2002). The cross section values of this paper are currently being revised;
 K.A. Snover, private communication.}
\bibitem{Stried}{F. Strieder {\it et al}., Nucl. Phys. {\bf A696}, 219 (2001).}
\bibitem{lagy} {L. T. Baby {\it et al}., Phys. Rev. Lett. {\bf 90}, 022501 
(2003).} 
\bibitem{weissn} {L. Weissman, M. Huyse, P. Van den Bergh and P. Van Duppen,
  Nucl. Instr. and Meth. {\bf A 452}, 147 (2000).}
\bibitem{Weiss1}{L. Weissman, M. Hass, V. Popov, Nucl. Instr. and  Meth.,  
A {\bf 400}, 409 (1999).}
\bibitem{Filip} {B.W. Filippone {\it et al}., Phys. Rev. {\bf C 28}, 2222 
(1983)}.
\bibitem{PSI} {U. K\"oster {\it et al}., EMIS-14, (2002); 
to be published in Nucl. Instr. Meth. B.}
\bibitem{hardy} {J.C. Hardy {\it et al}., Appl. Radiation and Isotopes, 
{\bf56}, 65 (2002).}
\bibitem{ndata} {S.Y.F. Chu, L.P. Ekstrom and R.B. Firestone:
 nucleardata.nuclear.lu.se/nucleardata/toi/.}
\bibitem{tuli} {Nuclear Wallet Cards, 6th Edition, 2000, Editor J.K. Tuli.}
\bibitem{war} {E.K. Warburton, Phys. Rev. {\bf C 33}, 303 (1986).}
\bibitem{ralph} {R.H. France {\it et al}., Nuclei in the Cosmos VII, 2002, 
Fuji-Yoshida, Japan.}
\bibitem{jeng} {B.K. Jennings, S. Karataglidis, and T.D. Shoppa, Phys. Rev. 
C 58, 3711 (1998).}
\bibitem{DB}{P. Descouvemont and D. Baye, Nucl. Phys. {\bf A567}, 341 (1994).}
\bibitem{bahc2}{J.N. Bahcall, M.H. Pinsonneault and S. Basu, Astrophys. J, 555,
990 (2001).}
\bibitem{bahc3}{J.N. Bahcall, arXiv:astro-ph/0209080.}


     \end{thebibliography}
\end{document}